\documentclass[aps,pra,twocolumn,amsmath,floatfix,showpacs,superscriptaddress]{revtex4}

\usepackage{graphicx}
\usepackage{color}
\usepackage{marvosym}
\usepackage{wasysym}
\usepackage{pifont}
\usepackage{amssymb}
\usepackage{amsmath}
\usepackage{amsfonts}
\usepackage{psfrag}
\usepackage[dvips]{epsfig}
\usepackage{epsfig}

\newcommand{\ud}{\mathrm{d}}

\newcommand{\br}{{\bf r}}

\newcommand{\bal}{\begin{align}}
\newcommand{\eal}{\end{align}}

\newcommand{\vv}[1]{\tilde{#1}}
\newcommand{\be}{\begin{equation}}
\newcommand{\ee}{\end{equation}}
\newcommand{\la}{\label}

\newcommand{\ie}{{\rm i.e.}}
\newcommand{\eg}{{\rm e.g.}}
\newcommand{\ba}{\begin{eqnarray}}
\newcommand{\ea}{\end{eqnarray}}
\newcommand{\inspa}{\Gamma_{\kappa}}

\begin{document}

\title{Core sizes and dynamical instabilities of giant vortices in dilute Bose-Einstein condensates}

\author{Pekko Kuopanportti}\email{pekko.kuopanportti@tkk.fi}\affiliation{Department of Applied Physics/COMP, Aalto University School of Science and Technology, P.\,O. Box 15100, FI-00076 AALTO, Finland}
\author{Emil Lundh}\affiliation{Department of Physics, Ume\aa~University, SE-90187 Ume\aa, Sweden}
\author{Jukka A. M. Huhtam\"aki}\affiliation{Department of Applied Physics/COMP, Aalto University School of Science and Technology, P.\,O. Box 15100, FI-00076 AALTO, Finland}\affiliation{Department of Physics, Okayama University, Okayama 700-8530, Japan}
\author{Ville Pietil\"a}\affiliation{Department of Applied Physics/COMP, Aalto University School of Science and Technology, P.\,O. Box 15100, FI-00076 AALTO, Finland}\affiliation{Australian Research Council Centre of Excellence for Quantum Computer Technology, School of Electrical Engineering and Telecommunications, University of New South Wales, Sydney NSW 2052, Australia}
\author{Mikko M\"ott\"onen}\affiliation{Department of Applied Physics/COMP, Aalto University School of Science and Technology, P.\,O. Box 15100, FI-00076 AALTO, Finland}\affiliation{Australian Research Council Centre of Excellence for Quantum Computer Technology, School of Electrical Engineering and Telecommunications, University of New South Wales, Sydney NSW 2052, Australia}\affiliation{Low Temperature Laboratory, Aalto University School of Science and Technology, P.\,O. Box 13500, FI-00076 AALTO, Finland}
\date{\today}

\begin{abstract}
Motivated by a recent demonstration of cyclic addition of quantized vorticity into a Bose-Einstein condensate, the vortex pump, we study dynamical instabilities and core sizes of giant vortices. The core size is found to increase roughly as a square-root function of the quantum number of the vortex, whereas the strength of the dynamical instability either saturates to a fairly low value or increases extremely slowly for large quantum numbers. Our studies suggest that giant vortices of very high angular momenta may be achieved by gradually increasing the operation frequency of the vortex pump.
\end{abstract}

\hspace{5mm}

\pacs{03.75.Kk, 03.75.Lm, 67.85.De}

\maketitle

\section{Introduction}\label{sc:intro} 

One of the signatures of Bose-Einstein condensation is the occurrence of superfluidity. In superfluids, particle currents can flow without dissipation, and the fluid is characterized by zero viscosity. A stable quantized vortex is a hallmark of such frictionless flow. These coherent whirlpools manifest the fundamental differences in the rotational characteristics of quantum and classical fluids.  The study of quantized vortices~\cite{Fetter2009} has flourished especially after their first realization in dilute Bose-Einstein condensates (BECs)~\cite{Matthews1999}. 

Large arrays of singly quantized vortices have been observed in rapidly rotating clouds of cold alkali metal atoms~\cite{Madison2001,Abo-Shaeer2001,Bretin2004,Stock2005}. Furthermore, multiquantum vortices, for which the phase of the condensate order parameter winds an integer multiple $\kappa$ of $2\pi$, have been created in rotating clouds by using an evaporative cooling technique to spin up the condensate~\cite{Haljan2001} in combination with a resonant laser beam focused at the center of the cloud~\cite{LosEngeles}. The giant vortices were reported to have surprisingly long lifetimes owing to strong Coriolis forces. However, the condensate states obtained in these experiments were not close to a pure, symmetric multiquantum vortex state, since the giant vortices were accompanied by several single-quantum vortices. Dynamics of such giant vortex states have since been investigated both experimentally and theoretically~\cite{Tapsa2004,Tapsa2005}.

Multiquantum vortices have also been created in an initially nonrotating cloud by transferring angular momentum into the condensate by a Laguerre-Gaussian laser beam~\cite{Andersen2006,Tapsa2008} and by a topological phase engineering method~\cite{Ville2008} which utilizes the spin degree of freedom of the condensate and its coupling to an external magnetic field~\cite{Nakahara2000,Isoshima2000,Ogawa2002,Mikko2002,Leanhardt2002,Leanhardt2003,Kumakura2006}. States with a multiquantum vortex are in general dynamically unstable: the state may annihilate due to a slight perturbation even in the absence of dissipation. Dynamical stability of multiquantum vortices has been investigated theoretically~\cite{multi1,Tapsa2002,multi2,multi3,multi4}, and splitting of multiquantum vortices into single-quantum ones has been studied both experimentally and theoretically~\cite{splitting1,splitting2,splitting3,splitting4,splitting5,splitting6,splitting7}. In addition, dynamical stability of coreless vortices~\cite{Ville2007,Takahashi2009} and vortex clusters~\cite{Crasovan2002,Crasovan2003,Mikko2005,Ville2006,Li2008,Seman2009} has been investigated.

Lately, it has been suggested that vortices with arbitrarily large winding numbers could be created by cyclically pumping vorticity into a BEC~\cite{pumppu}. The technique is based on the topological phase engineering method: the spin degree of freedom of the condensate is controlled locally by alternating external quadrupole and hexapole magnetic fields, and a fixed amount of vorticity is added into the system in each cycle. In the simulations of Ref.~\cite{pumppu}, the operation of the vortex pump was demonstrated both adiabatically, in which case an additional optical potential is required to confine the condensate during the pumping cycle, and nonadiabatically without any optical potentials, but with the cost of losing a part of the atoms from the trap. Later, the vortex pump was investigated numerically using only a single multipole magnetic field, albeit an additional uniform field was introduced~\cite{Xu2008}. This scheme has the advantage that no added vorticity is removed during the cycle, but it necessarily requires an optical potential alongside the magnetic fields. Recently, it has also been theoretically demonstrated that topological phase engineering can be used to create a Dirac monopole in a spinor BEC~\cite{Monopoli2009}.

In this paper, we study the vortex core sizes and dynamical instabilities of axisymmetric giant vortex states in dilute nonrotated BECs as functions of the vorticity quantum number and the effective atom-atom interaction strength. We restrict our investigation to the zero-temperature limit and neglect the possible additional effects due to thermal atoms. Hence, the analysis presented here provides an upper bound for the lifetime of the giant vortices.

Our investigation is motivated by two aspects. Firstly, the studies of multiquantum vortices in dilute BECs to date have focused on the regime $\kappa \leq 5$, probably because previously there has been no practically realizable techniques available to create isolated vortices with large quantum numbers in a controlled manner. Now, however, the vortex pump provides such an opportunity. Secondly, we aim to determine how large winding numbers the pump can reach for a single vortex. In principle, vortex pumping enables a controlled production of vortices with arbitrarily large quantum numbers. However, in real experiments the quantum number will be limited by the stability properties of the vortex states. In particular, the existence of dynamical instabilities renders the multiquantum vortices prone to splitting into single-quantum vortices. The frequencies of these instabilities can be used to approximate the lifetimes of the multiquantum vortex states. On the other hand, the core size of the vortex is known to increase with increasing winding number. As discussed in Sec.~\ref{sc:results}, this enables us to speed up the pumping cycle as the vorticity of the condensate increases. Hence, the interplay between the core size and the lifetime of the vortex as a function of the winding number ultimately determines the maximum winding number reachable with the vortex pump.

The remainder of this paper is organized as follows. We present our theoretical model in Sec.~\ref{sc:theory} and also consider the Thomas-Fermi (TF) approximation for giant vortex states. In Sec.~\ref{sc:nonint}, we formulate the analytical approximation in the noninteracting limit. Section~\ref{sc:results} is devoted to presenting our numerical results on the core sizes and dynamical instabilities of multiquantum vortices. We compare the numerically obtained core sizes with the analytical results of the TF approximation and consider how well the noninteracting limit can be used to describe the dynamical instabilities in the presence of finite atom-atom interactions. In Sec.~\ref{sc:discussion}, we summarize and discuss the main results of our study.

\section{Theoretical framework}\label{sc:theory}

In the zero-temperature limit, the complex-valued order parameter field $\Psi(\br,t)$ of a dilute BEC is described by the time-dependent Gross-Pitaevskii (GP) equation,
\begin{equation}\label{eq:tGPE}
i\hbar\frac{\partial}{\partial t}\Psi(\br,t) = \left[ {\cal H}+ g|\Psi(\br,t)|^2 \right] \Psi(\br,t),
\end{equation}
where ${\cal H}$ denotes the single-particle Hamiltonian and the atom-atom interaction strength parameter $g$ is related to the vacuum $s$-wave scattering length $a$ and the atomic mass $m$ by $g=4\pi\hbar^2 a / m$ \cite{Pethick2002}. For BECs enclosed in nonrotated cylindrically symmetric harmonic traps, the Hamiltonian ${\cal H}$ is expressed in the cylindrical coordinates $(r,\phi,z)$ as
\begin{equation}\label{eq:hamiltonian}
{\cal H} = -\frac{\hbar^2}{2m}\nabla^2 + \frac{1}{2}m\omega_r^2\left(r^2+\lambda^2 z^2\right),
\end{equation}
where $\omega_r$ is the harmonic oscillator frequency in the radial direction and $\lambda = \omega_z / \omega_r$ defines the aspect ratio of the trap. The order parameter is normalized according to $\int |\Psi(\br)|^2 \ud \br = N$, with $N$ being the total number of particles.

Stationary states of the system are solutions to the time-independent GP equation,
\begin{equation}\label{eq:GPE}
\left[ {\cal H}+ g|\Psi(\br)|^2 -\mu \right] \Psi(\br) = 0,
\end{equation}
where $\mu$ denotes the chemical potential. In order to study the small-amplitude oscillations about a given stationary state, we write the order parameter in the form $\Psi(\br,t)=\exp(-i\mu t/\hbar)\left[\Psi(\br) +\chi(\br,t)\right]$, where $\chi(\br,t)$ is assumed to have a small $L^2$-norm compared with $\Psi(\br)$. By substituting this trial function into Eq.~(\ref{eq:tGPE}), using the decomposition
\be\label{eq:oscillation}
\chi(\br,t)=\sum_q \left[ u_q(\br) e^{-i\omega_q t} + v_q^\ast(\br)e^{i\omega_q^\ast t} \right],
\ee
and linearizing with respect to $\chi(\br,t)$, we arrive at the Bogoliubov equations
\begin{equation}\label{eq:bogo}
\left( \begin{array}{cc} {\cal L} & g\Psi^2 \\ -g [\Psi^\ast]^2 & -{\cal L} \end{array} \right) \left( \begin{array}{c} u_q(\br) \\ v_q(\br) \end{array} \right) = \hbar \omega_q \left( \begin{array}{c} u_q(\br) \\ v_q(\br) \end{array} \right),
\ee
where we have denoted ${\cal L} = {\cal H} + 2g|\Psi|^2 - \mu$. An alternative derivation utilizing a canonical transformation of second-quantized operators shows that the functions $u_q(\br)$ and $v_q(\br)$ are the quasiparticle amplitudes and $\omega_q$ is the eigenfrequency of the elementary excitation mode corresponding to the index $q$ \cite{Pethick2002}.

The Bogoliubov equations can be used for the stability analysis of the system. If the quasiparticle spectrum $\{\omega_q\}$ contains excitations with a positive norm $\int\left[ |u_q|^2-|v_q|^2 \right]\ud\br$ but a negative eigenfrequency $\omega_q$, the corresponding stationary state is energetically unstable. On the other hand, the stationary state is dynamically unstable if the quasiparticle spectrum contains at least one eigenfrequency with a positive imaginary part~\cite{multi1}. As can be observed from Eq.~(\ref{eq:oscillation}), the occupations of such complex-frequency modes initially increase exponentially in time, and consequently small perturbations of a dynamically unstable stationary state typically lead to large changes in its structure. In particular, dynamically unstable multiquantum vortex states are unstable against splitting of the vortex, and for such states the quantity $\max_q[|\mathrm{Im}(\omega_q)|]$ can be used to estimate the splitting tendency of the vortex~\cite{splitting2}. However, the dynamically unstable modes quickly drive the system beyond the linear regime of Eq.~(\ref{eq:oscillation}), and hence the dynamics must instead be described with the time-dependent GP equation.

In this paper, we study the properties of stationary multiquantum vortex states as functions of the winding number $\kappa$ of the vortex. Since the external potential in Eq.~(\ref{eq:hamiltonian}) is rotationally symmetric, a stationary state containing a $\kappa$-quantum vortex can be written in the form
\be\label{eq:psi}
\Psi(r,\phi,z) = \sqrt{n(r,z)}e^{i\kappa\phi},
\ee
where $n(r,z)$ is the particle density of the condensate. In order to simplify the analysis, we limit our considerations to pancake-shaped BECs, \ie, we assume that the harmonic confinement in the axial direction is sufficiently tight such that the axial harmonic oscillator length $a_z = \sqrt{\hbar/m\omega_z}$ is much smaller than the characteristic length scale of density variations in the radial direction. This approximation is accurate for $\lambda \gg 1$, and in such a case the $z$ dependence of the order parameter can be factored out as $\sqrt{n(r,z)}=f(r)\zeta_0(z)$, where $\zeta_0(z) = \exp[-z^2/(2a_z^2)]/\sqrt[4]{\pi a_z^2}$. Moreover, for this kind of stationary states, the azimuthal and axial dependences can be separated from the solutions of the Bogoliubov equations (\ref{eq:bogo}), and consequently the quasiparticle amplitudes can be expressed in the form 
\ba
u_q(r,\phi,z)&=&u_q(r)\zeta_0(z)e^{i(l_q+\kappa)\phi}, \nonumber \\ 
v_q(r,\phi,z)&=&v_q(r)\zeta_0(z)e^{i(l_q-\kappa)\phi}, \label{eq:qp_amplitudes}
\ea
where $l_q$ is an integer that determines the angular momentum of the quasiparticle excitation with respect to the condensate.

It is convenient to express all quantities using dimensionless variables, which are henceforth denoted with a tilde. Therefore, we measure length in the units of the harmonic oscillator length $a_r=\sqrt{\hbar/m\omega_r}$ and energy in units of $\hbar\omega_r$. Using the factorization of Eq.~(\ref{eq:psi}) with $\sqrt{n(r,z)}=f(r)\zeta_0(z)$, we arrive at the one-dimensional GP equation
\be\label{eq:dimless_GPE}
\left[\vv{\cal H}_\kappa+\vv{g}\vv{f}^2(\vv{r})- \vv{\mu} \right] \vv{f}(\vv{r})=0,
\ee
where the dimensionless interaction strength parameter is defined as $\vv{g}=2\sqrt{2\pi}Na/a_z$, the chemical potential $\vv{\mu}=\mu/(\hbar\omega_r)-\omega_z/(2\omega_r)$ contains the shift due to the $z$ displacement, and the dimensionless single-particle Hamiltonian is given by
\be
\vv{\cal H}_\kappa = -\frac{1}{2}\left(\frac{1}{\vv{r}}\frac{\partial}{\partial\vv{r}} + \frac{\partial^2}{\partial\vv{r}^2} - \frac{\kappa^2}{\vv{r}^2} - \vv{r}^2 \right).
\ee
Function $\vv{f}$ is normalized according to the condition $2\pi\int\vv{n}(\vv{r})\vv{r}\ud\vv{r}=1$, where $\vv{n}(\vv{r})=\vv{f}^2(\vv{r})$ is the areal probability density of the condensed particles. The dimensionless Bogoliubov equations for a $\kappa$-quantum vortex state are given by 
\be\label{eq:dimless_bogo}
\left( \begin{array}{cc} \vv{{\cal{L}}}_{l_q+\kappa} & \vv{g} \vv{f}^2(\vv{r})  \\ -\vv{g} \vv{f}^2(\vv{r}) & -\vv{{\cal{L}}}_{l_q-\kappa}\end{array} \right) \left( \begin{array}{c} \vv{u}_q(\vv{r}) \\ \vv{v}_q(\vv{r}) \end{array} \right) = \vv{\omega}_q \left( \begin{array}{c} \vv{u}_q(\vv{r}) \\ \vv{v}_q(\vv{r}) \end{array} \right),
\ee
where $\vv{{\cal{L}}}_{\kappa} = \vv{\cal H}_\kappa + 2 \tilde{g}\vv{f}^2(\tilde{r})-\vv{\mu}$. Using the values $a = 2.75\textrm{ nm}$ and $m=3.81 \times 10^{-26}\textrm{ kg}$ corresponding to ${}^{23}$Na atoms~\cite{Pethick2002} and taking the radial oscillator frequency and the trap asymmetry to be $\omega_r = 2\pi \times 20\textrm{ Hz}$ and $\lambda = 50$, which are typical values for traps used in experiments~\cite{Clade2009}, we obtain $a_r \approx 4.7\ \mu\mathrm{m}$ and $\vv{g}\approx 0.021 \times N$. Thus, for a ${}^{23}$Na condensate with $N \approx 10^4$ atoms, the effective interaction strength is of the order of $\vv{g} \approx 200$. 

For sufficiently large numbers of atoms, \ie, for large enough values of $\vv{g}$, a plausible description for the condensate order parameter is obtained by solving the GP equation in the Thomas-Fermi (TF) approximation~\cite{TF}. In our case, the TF solution corresponds to neglecting the radial derivatives of $\vv{f}(\vv{r})$ in Eq.~(\ref{eq:dimless_GPE}), which yields
\be\label{eq:TF-density}
\vv{n}_\mathrm{TF}(\vv{r}) = \frac{1}{\vv{g}}\left( \vv{\mu}_\mathrm{TF} - \frac{1}{2}\vv{r}^2 - \frac{1}{2}\frac{\kappa^2}{\vv{r}^2} \right),
\ee
if $\vv{n}_\mathrm{TF}(\vv{r}) > 0$, and $0$ otherwise. The inner and outer TF radii, between which $\vv{n}_\mathrm{TF}(\vv{r})$ is nonvanishing, are given by $\vv{r}_\mp^2=\vv{\mu}_\mathrm{TF}\mp\sqrt{\vv{\mu}_\mathrm{TF}^2-\kappa^2}$. The TF expression for the chemical potential, $\vv{\mu}_\mathrm{TF}$, is determined by the normalization condition $2\pi\int\vv{n}_\mathrm{TF}(\vv{r})\vv{r}\ud\vv{r}=1$, which yields the equation
\be\la{eq:TFapprox}
\frac{\vv{g}}{\pi} = \vv{\mu}_\mathrm{TF}^2 \left[ \theta-\frac{1-\theta^2}{2 }\ln \left(\frac{1-\theta}{1+\theta} \right) \right] \approx \vv{\mu}_\mathrm{TF}^2\theta^4, 
\ee
where $\theta = \sqrt{1-\kappa^2/\vv{\mu}_\mathrm{TF}^2}$. In Eq.~(\ref{eq:TFapprox}), we have approximated the bracketed function by $\theta^4$, which is accurate for $1/\sqrt{2} < \theta \leq 1$ (the relative error is less than 7\%). Finally, solving for the chemical potential gives
\be\label{eq:TF-mu}
\vv{\mu}_\mathrm{TF}= \sqrt{\frac{\vv{g}}{4\pi}} + \sqrt{\frac{\vv{g}}{4\pi}+\kappa^2},
\ee
and thus the approximation in Eq.~(\ref{eq:TFapprox}) is justified if $\vv{g} > \kappa^2$~\cite{remark1}. 

\section{Noninteracting limit}\label{sc:nonint}

In the limit of weak interactions, the instabilities can be calculated by treating the interaction strength $g$ as a perturbative parameter~\cite{multi4,kavoulakis2000}. The unperturbed wavefunctions are the single-particle eigenfunctions without radial nodes, 
\be
\vv{\varphi}_n(\tilde{r}) = \frac{1}{\sqrt{\pi |n|!}} \tilde{r}^{|n|} e^{-\tilde{r}^2/2},
\ee
where $n\in\mathbb{Z}$ and the angular dependence $\exp(i n\phi)$ has been factored out. Thus, for a $\kappa$-quantum vortex state, we have
\be
\tilde{f}(\tilde{r}) = \vv{\varphi}_{\kappa}(\tilde{r}),
\ee
and the corresponding Bogoliubov eigenstates are the particle states 
\be
\left(\begin{array}{c} \vv{u}_q^0(\tilde{r}) \\ \vv{v}_q^0(\tilde{r}) \end{array}\right) = \left(\begin{array}{c} \vv{\varphi}_{\kappa+l_q}(\tilde{r})\\
0 \end{array}\right),
\ee
with energies $\tilde\omega_q^0=|\kappa+l_q|-\kappa$, and the hole states 
\be \left(\begin{array}{c} \vv{u}_q^0(\tilde{r}) \\ \vv{v}_q^0(\tilde{r}) \end{array}\right) = \left(\begin{array}{c} 0\\ \vv{\varphi}_{\kappa-l_q}(\tilde{r})
\end{array}\right),
\ee
with energies $\tilde\omega_q^0=-|\kappa-l_q|+\kappa$. We note that for $l_q \leq \kappa$, the particle and hole states with equal angular momentum quantum numbers $l_q$ are degenerate. By applying degenerate-state perturbation theory, we find the Bogoliubov energies 
\be\label{pertenergy}
\tilde\omega_q = l_q + \vv{g}\frac{I_{+}-I_{-}}{2} \pm \vv{g}\sqrt{\left(I_{+}+I_{-}-I_{c}\right)^2 - |I_{x}|^2},
\ee
where 
\ba
I_{\pm} &=& 2\pi \int\vv{\varphi}_{\kappa}(\tilde{r})^2\vv{\varphi}_{\kappa\pm l_q}(\tilde{r})^2 \tilde{r}\,\ud\tilde{r}, \\
I_{c} &=& 2\pi \int \vv{\varphi}_{\kappa}(\tilde{r})^4 \tilde{r}\,\ud\tilde{r} ,
\ea
and
\ba
I_{x} &=& 2\pi \int \vv{\varphi}_{\kappa}(\tilde{r})^2 \vv{\varphi}_{\kappa + l_q}(\tilde{r}) \vv{\varphi}_{\kappa - l_q}(\tilde{r}) \tilde{r}\,\ud\tilde{r}. 
\ea
The integrals can be evaluated using the formula~\cite{kavoulakis2000}
\ba
2\pi \int\varphi_{m}(\tilde{r})\varphi_{n}(\tilde{r}) \varphi_{k}(\tilde{r})\varphi_{l}(\tilde{r})\tilde{r}\,\ud\tilde{r}= \nonumber \\
\frac{1}{2\pi} \frac{[(|m|+|n|+|k|+|l|)/2]!}{2^{(|m|+|n|+|k|+|l|)/2} \sqrt{|m|! |n|! |k|! |l|!}}.
\ea

We are interested in the maximum imaginary part of the Bogoliubov eigenfrequencies. In the noninteracting limit, it is found by minimizing the expression inside the square root in Eq.~(\ref{pertenergy}). Note that the perturbation theory does not describe the eventual disappearance of a given unstable mode at higher values of $\vv{g}$, but incorrectly predicts that the instabilities grow linearly with $\vv{g}$. Nevertheless, for a given winding number $\kappa$, one can determine the angular momentum quantum number $l_{\mathrm{dom}}^0$ that gives the dominant instability in the noninteracting limit. These analytical results are presented and compared to the numerical ones in Sec.~\ref{sc:results}. 

\section{Results}\label{sc:results}

We have solved the stationary states and the corresponding Bogoliubov excitation spectra, Eqs.~(\ref{eq:dimless_GPE}) and~(\ref{eq:dimless_bogo}), for the parameter ranges $0\leq\kappa\leq100$ and $0\leq\vv{g}\leq10^4$. In the calculations, we have used finite difference methods with grid sizes of roughly 200 points and {\footnotesize LAPACK} numerical library for solving the discretized eigenvalue problem. 

In order to illustrate a typical profile of a giant vortex state and its TF approximation, Fig.~\ref{fig:wfs_g1000} shows the modulus of the order parameter for $\vv{g}=1000$ and different values of $\kappa$, together with the corresponding TF profiles. As expected, the TF approximation fails near the inner and outer surfaces of the condensate. Inside the cloud, the TF profile provides an accurate estimate as long as the condition $\vv{g}>\kappa^2$ is satisfied. The state with $\kappa=40$ does not satisfy this criterion, and consequently the TF profile significantly deviates from the numerical solution also away from the surface of the cloud.

\subsection{Size of the vortex core}\label{subsc:core}

Here, we study the radius of the vortex core, $\vv{r}_\mathrm{c}$, as a function of the winding number $\kappa$ and the interaction strength $\vv{g}$. We define the radius as the smallest solution of the equation
\be\label{eq:rc}
\vv{f}^2(\vv{r}_\mathrm{c}) = \frac{3}{4}\max_{\vv{r}} \vv{f}^2(\vv{r}),
\ee
where $\vv{f}^2(\vv{r})$ is the areal probability density of the stationary $\kappa$-quantum vortex state satisfying Eq.~(\ref{eq:dimless_GPE}). Our investigation of the vortex core size is partly motivated by the fact that for the vortex pump~\cite{pumppu}, the core radius essentially determines the maximum adiabatic pumping speed in the following way. A characteristic energy associated with the magnetic field is given by the minimum energy separation $\epsilon_\textrm{mag}=g_\mathrm{F}\mu_\mathrm{B} B_\mathrm{min}$, where $g_\mathrm{F}$ is the Land\'{e} $g$ factor, $\mu_\mathrm{B}$ is the Bohr magneton, and $B_\mathrm{min}$ is the minimum magnetic field strength in the condensate region during the bias field inversion. The field strength $B$ is minimized when the spatially homogeneous bias field $B_z$ crosses zero, at which point the magnitude of the field increases linearly with distance from the $z$ axis. Thus, the relevant minimum field strength in the presence of a vortex becomes $B_\mathrm{min}=B_\perp'r_\mathrm{c}$, where $B_\perp'$ denotes the gradient of the perpendicular multipole field. To guarantee adiabaticity, the time derivative of the Hamiltonian should be small compared with the energy separation of its instantaneous eigenstates. In our case, this roughly means that the bias field $B_z$ should be inverted such that
\be
\frac{\hbar g_\mathrm{F}\mu_\mathrm{B}}{\epsilon_\textrm{mag}^2} \times \frac{\partial B_z}{\partial t} = \frac{\hbar}{g_\mathrm{F}\mu_\mathrm{B} (B_\perp')^2 r_\mathrm{c}^2}\times \frac{\partial B_z}{\partial t} \ll 1.
\ee
Since the core radius $r_\mathrm{c}$ is observed to increase with $\kappa$, it is possible to speed up the pumping cycle without challenging adiabaticity as more vortices are pumped into the BEC.

In order to obtain an analytical approximation for the vortex core size, we use the TF density profile, Eq.~(\ref{eq:TF-density}). Inserting Eq.~(\ref{eq:TF-density}) into Eq.~(\ref{eq:rc}), we find the expression
\be\label{eq:TF-core}
\vv{r}_\mathrm{c}^\mathrm{TF}=\frac{1}{2}\sqrt{\vv{\mu}_\mathrm{TF}+3\kappa - \sqrt{\vv{\mu}_\mathrm{TF}^2+6\kappa\vv{\mu}_\mathrm{TF}-7\kappa^2}},
\ee
which, with the help of Eq.~(\ref{eq:TF-mu}), can be directly compared with the numerical results. Furthermore, since the TF inner radius $\vv{r}_-$ gives an estimate for the radius below which there are no particles, it could also be used to evaluate the adiabaticity condition of the vortex pump as discussed above. 

On the other hand, the core size of single-quantum vortices is typically approximated by the healing length $\xi=1/\sqrt{8\pi\bar{n}a}$, which describes the characteristic distance over which the condensate density tends to its bulk value $\bar{n}$ when subjected to a localized perturbation in an otherwise homogeneous system~\cite{Pethick2002}. For multiquantum vortices, we can use a similar approach to obtain a simple approximation for the core radius. However, since the size of the giant vortex is comparable to that of the whole condensate, the inhomogeneity of the trapped gas can no longer be neglected, but instead the trapping potential term must be included in the energy balance. Hence, we equate the kinetic energy with the potential and interaction energy terms at $\vv{r}=\vv{\xi}_\mathrm{c}$ and obtain
\be
\frac{\kappa^2}{2\vv{\xi}_\mathrm{c}^2}=\frac{1}{2}\vv{\xi}_\mathrm{c}^2 + \vv{U},
\ee
where $\vv{U}$ denotes the interparticle interaction. For a vortex-free condensate in a harmonic potential in the TF limit, one finds that near the center of the cloud $\vv{U}\propto N^{2/5}$~\cite{Pethick2002}. Since $\tilde{g}\propto N$, we make the crude estimate $\vv{U} \approx \vv{g}^{2/5}$ to obtain the core radius approximation 
\be\label{eq:xi_c}
\vv{\xi}_\mathrm{c} = \sqrt{\sqrt{\kappa^2 + \vv{g}^{4/5}}-\vv{g}^{2/5}},
\ee
which should be compared with the more rigorous approximation given by Eqs.~(\ref{eq:TF-core}) and~(\ref{eq:TF-mu}). We note that for large quantum numbers $\kappa \gg \tilde{g}^{2/5}$, Eqs.~\eqref{eq:TF-core} and~\eqref{eq:xi_c} predict that the core radius increases as a square root of $\kappa$.

In Fig.~\ref{fig:cores}, we plot the vortex core radius as a function of $\kappa$ for $\vv{g}=1$ and $10^4$ together with the approximations of Eqs.~(\ref{eq:TF-core}) and~(\ref{eq:xi_c}). Figure~\ref{fig:cores_g} shows the radius versus the interaction strength $\vv{g}$ for different winding numbers. As expected, $\vv{r}_\mathrm{c}$ monotonously increases with increasing $\kappa$ and decreases with increasing $\vv{g}$.  As a function of $\kappa$, the asymptotic behavior $\vv{r}_\mathrm{c}\propto\sqrt{\kappa}$ is observed for $\kappa^2\gg \vv{g}$, as predicted by Eqs.~(\ref{eq:TF-core}) and~(\ref{eq:xi_c}). The reduction of the core size with the interaction strength is relatively slow, and $\vv{r}_\mathrm{c}$ decreases by a few tens of percent over the range $0 < \vv{g} \leq 10^4$. In general, the analytical TF expression for the vortex core radius, Eq.~(\ref{eq:TF-core}), yields good agreement with the numerical results, even for small values of $\vv{g}$. On the other hand, the simpler approximation of Eq.~(\ref{eq:xi_c}) is observed to give accurate results for sufficiently small values of $\vv{g}$ and large values of $\kappa$.

\begin{figure}[tb]
\includegraphics[width=210pt]{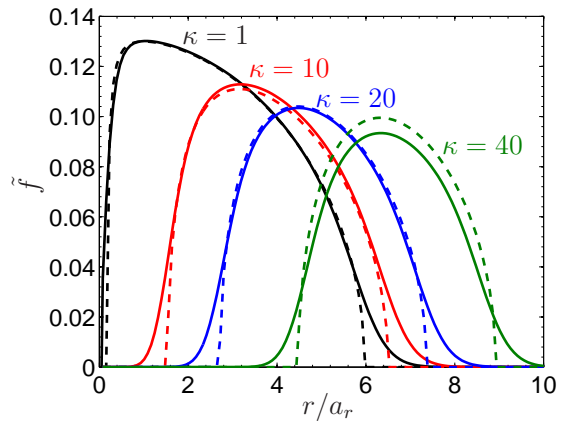}
\caption{(Color online) Numerical solution (solid curve) and the TF approximation (dashed curve) for $\vv{f}(\vv{r})$ as a function of $\vv{r}$ for $\vv{g}=1000$. The curves correspond to the winding numbers $\kappa=1,10,20,40$ as indicated.}
\label{fig:wfs_g1000}
\end{figure}

\begin{figure}[tb]
\includegraphics[width=210pt]{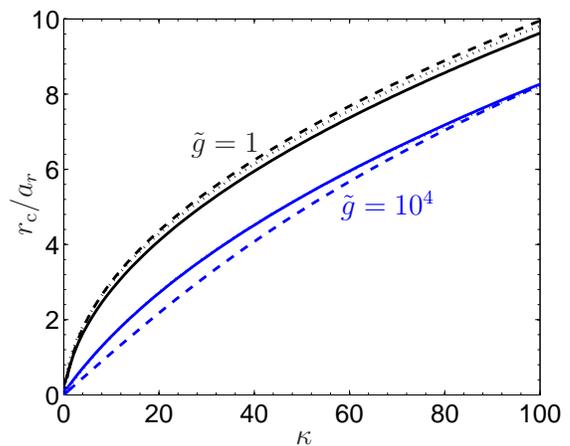}
\caption{(Color online) Radius of the vortex core as a function of the winding number $\kappa$ for $\vv{g}=1$ and $\vv{g}=10^4$. The solid curve shows the numerically obtained radius $\vv{r}_\mathrm{c}$, the dotted curve is the TF approximation $\vv{r}_\mathrm{c}^\mathrm{TF}$ given by Eqs.~(\ref{eq:TF-mu}) and~(\ref{eq:TF-core}), and the dashed curve corresponds to the approximation $\vv{\xi}_\mathrm{c}$, Eq.~(\ref{eq:xi_c}). For $\vv{g}=10^4$, the solid curve lies on top of the dotted curve.}
\label{fig:cores}
\end{figure}

\begin{figure}[tb]
\includegraphics[width=210pt]{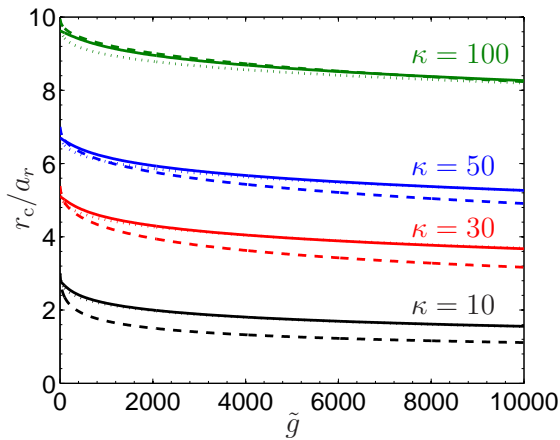}
\caption{(Color online) Radius of the vortex core as a function of the interaction strength $\vv{g}$. The solid curve shows the numerically obtained radius $\vv{r}_\mathrm{c}$, the dotted curve is the TF approximation $\vv{r}_\mathrm{c}^\mathrm{TF}$, and the dashed curve corresponds to the approximation $\vv{\xi}_\mathrm{c}$. The curves correspond to the winding numbers $\kappa =10,30,50,100$ as indicated.}
\label{fig:cores_g}
\end{figure}

\subsection{Dynamical instabilities}\label{subsc:instabilities}

Next, we investigate the dynamical instability properties of the vortex states as functions of the winding number and the interaction strength. It is known from earlier numerical simulations that in pancake-shaped condensates, the multiquantum vortex can be dynamically stable or unstable depending on the interaction strength $\vv{g}$~\cite{multi1,multi2}. In a harmonic trap, the two-quantum vortex state can have a dynamical instability mode only for the angular momentum quantum number $|l_q|=2$. A vortex with a higher winding number $\kappa$ can have complex-frequency modes also at larger values of $|l_q|$, each mode corresponding to an $|l_q|$-fold symmetric splitting pattern \cite{splitting6}. These observations are also predicted by the noninteracting approximation of Sec.~\ref{sc:nonint}, according to which degenerate modes exist for those $l_q$ that satisfy $l_q\leq\kappa$. Because the number of different $l_q$-modes supporting complex eigenfrequencies tends to increase with $\kappa$, we expect the multiquantum vortices to generally become more and more unstable as the vorticity increases.

Since we are interested in the general degree of dynamical instability of a vortex state with a given winding number $\kappa$, we need a quantity that is independent of the interaction strength $\vv{g}$. Therefore, we define the dimensionless instability parameter $\inspa$ by
\begin{equation}\label{eq:inspa}
\inspa = \max_{\vv{g}} \left\{ \max_q \left[ | \mathrm{Im} (\vv{\omega}_q ) | \right] \right\} .
\end{equation}
In other words, we solve Eq.~(\ref{eq:dimless_bogo}) for fixed values of $\kappa$ and $\vv{g}$ and maximize the quantity $|\mathrm{Im}(\vv{\omega}_q)|$ over each eigenfrequency spectrum, after which we vary the interaction strength and maximize $\max_q [|\mathrm{Im}(\vv{\omega}_q)|]$ as a function of $\vv{g}$. The parameter $\inspa$ provides a measure for the splitting tendency of a $\kappa$-quantum vortex, and $1/\inspa$ can be used to estimate the lifetime of the vortex state.  Furthermore, it was found in Ref.~\cite{splitting2} that a highly elongated BEC, \ie, one with a small aspect ratio $\lambda$, can be described by a local-density approximation in which the condensate is viewed as a stack of two-dimensional systems with different values of $\vv{g}$. If the conditions for two-dimensional instability are met locally at a point $z$, the vortex splitting will commence at that point and subsequently migrate along the vortex line. Thus, finding the maximum of the quantity $\max_q [|\mathrm{Im}(\vv{\omega}_q)|]$ with respect to $\vv{g}$ is equivalent to finding the dominant unstable mode of a cigar-shaped BEC in this approximation.

The numerically obtained values of $\inspa$ are presented in Fig.~\ref{fig:maxim}. Figure~\ref{fig:gmax} shows the values of the interaction strength, $\vv{g}_\mathrm{dom}$, for which the dominant complex-frequency mode occurs. The angular momentum quantum number $l_\mathrm{dom}$ of the dominant mode is also indicated. As expected, $\inspa$ is a strictly increasing function of $\kappa$. However, we observe that the value of $\inspa$ either saturates to a relatively low value of about 0.3 or increases very slowly for large winding numbers. In addition, the instability parameter exhibits quasiperiodic behavior that results from the stepwise increments in the value of $l_\mathrm{dom}$. Quasiperiodic behavior is observed also for $\vv{g}_\mathrm{dom}$ in Fig.~\ref{fig:gmax}: $\vv{g}_\mathrm{dom}$ increases within regions of constant $l_\mathrm{dom}$ but exhibits a sudden drop each time $l_\mathrm{dom}$ increases.

\begin{figure}[tb]
\includegraphics[width=210pt]{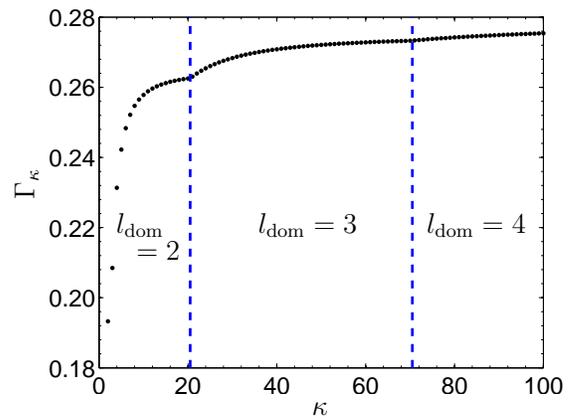}
\caption{(Color online) Dimensionless dynamical instability parameter $\inspa$ of a giant vortex state as a function of its winding number $\kappa$. The dashed vertical lines separate regions where the angular momentum quantum numbers of the dominant complex-frequency modes, $l_\mathrm{dom}$, are different.}
\label{fig:maxim}
\end{figure}

\begin{figure}[tb]
\includegraphics[width=210pt]{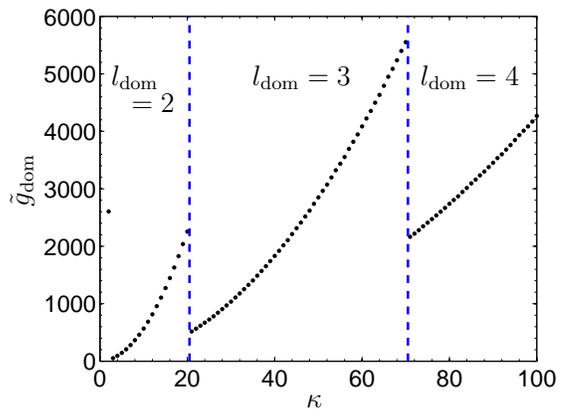}
\caption{(Color online) Values of the dimensionless interaction strength $\tilde{g}$ at which the largest dynamical instabilities occur. The angular momentum quantum number of the dominant complex-frequency mode, $l_\mathrm{dom}$, is also indicated.}
\label{fig:gmax}
\end{figure}

In earlier studies for $2 \leq\kappa\leq 5$, the dominant complex-frequency mode over different values of the interaction strength was found at the angular momentum quantum number $l_q=2$. According to our results, the $l_q=2$ mode dominates up to the winding number $\kappa=20$, after which the largest imaginary part is provided by an excitation mode with $l_q=3$. The subsequent transition to $l_\mathrm{dom}=4$ occurs at $\kappa=70$. These findings should be compared with the analytical results, displayed in Fig.~\ref{fig:nonint}, for the angular momentum quantum number $l_{\mathrm{dom}}^0$ that provides the largest complex-frequency mode in the noninteracting limit for a given winding number. We have also indicated the numerically obtained value of $\vv{g}$ above which the angular momentum quantum number of the most unstable mode becomes different from $l_{\mathrm{dom}}^0$. We observe that $l_{\mathrm{dom}}^0$ increases slowly as a function of $\kappa$. For small $\kappa$, the most unstable mode has the quantum number $l_{\mathrm{dom}}^0$ for exactly $l_{\mathrm{dom}}^0$ consecutive values of $\kappa$. However, this numerical coincidence ceases to hold after $l_{\mathrm{dom}}^0=7$. We observe that in the noninteracting limit, the dominant complex-frequency modes generally have much larger angular momentum quantum numbers than is found numerically for unrestricted $\vv{g}$. Moreover, the range of validity for the noninteracting result is restricted to very small values of $\vv{g}$.

\begin{figure}[tb]
\includegraphics[width=210pt]{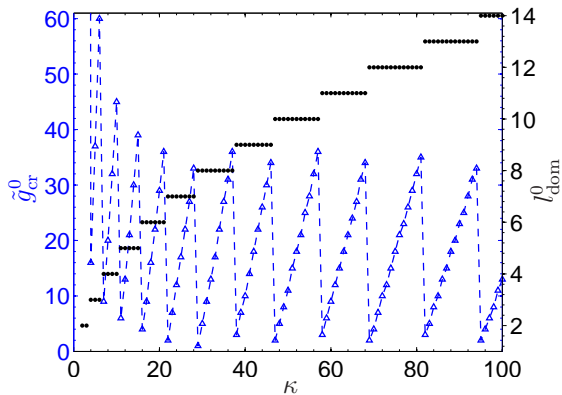}
\caption{(Color online) Dots show the angular momentum quantum numbers $l_{\mathrm{dom}}^0$ yielding the largest dynamical instability in the noninteracting limit. The triangles denote the critical interaction strength, $\tilde{g}_{\mathrm{cr}}^0$, above which the noninteracting result for $l_\mathrm{dom}$ fails. The dashed line is a guide to the eye.}
\label{fig:nonint}
\end{figure}

\section{Discussion}\label{sc:discussion}

In conclusion, we have studied the core sizes and dynamical instabilities of multiquantum vortices in pancake-shaped nonrotated BECs as a function of the winding number $\kappa$. We found the core size to be a strictly increasing function of the winding number with the asymptotic behavior $\vv{r}_\mathrm{c}\propto \sqrt{\kappa}$ for large values of $\kappa$. The dependence of $\vv{r}_\mathrm{c}$ on the interaction strength $\vv{g}$ turned out to be weak. On the other hand, the maximum strength of the dynamical instability of a $\kappa$-quantum vortex state was observed to increase slowly with $\kappa$ or even to saturate to a relatively low value. Based on these results, the time scales for the splitting of giant vortices are expected to decrease only slowly with increasing winding number. In addition, we found that quasiparticle excitations with the angular momentum quantum number $l_q=2$ dominate the instability spectra up to the winding number $\kappa=20$, after which consequent transitions to higher modes are observed.

We found that the analytical TF approximation provides good estimates for the exact particle densities as long as the condition $\vv{g} > \kappa^2$ is satisfied. Moreover, the TF estimate for the radius of the multiquantum vortex was found to be in good agreement with the numerically determined radius for all parameter values $\{\kappa,\vv{g}\}$ considered. 

From the point of view of the vortex pump~\cite{pumppu}, our results are encouraging. Qualitatively, the efficiency of the pump in producing vortices with large winding numbers is described by the ratio $\tau_\mathrm{sp}/T_\mathrm{inv}$, where $\tau_\mathrm{sp}$ is a characteristic time scale for vortex splitting and $T_\mathrm{inv}$ is the bias field inversion time. Since the radius of the vortex core increases with the winding number approximately as $r_\mathrm{c}\propto\sqrt{\kappa}$, it should be possible to gradually decrease $T_\mathrm{inv}$ and still retain the adiabaticity of the process. On the other hand, we expect the splitting times $\tau_\mathrm{sp}$ to decrease only slowly with $\kappa$ for large vorticities. Consequently, it should be possible to actually increase the ratio $\tau_\mathrm{sp}/T_\mathrm{inv}$ after a sufficient amount of vorticity has accumulated into the condensate. Hence, our results suggest that giant vortices with very high quantum numbers may be achieved by gradually increasing the operation frequency of the pump, \eg, such that $T_\mathrm{inv}\propto \kappa^{-1}$.

\begin{acknowledgments}
The Academy of Finland and the Emil Aaltonen foundation are acknowledged for financial support. J.~H. thanks JSPS for support. J.~H. and P.~K. acknowledge support from the V\"ais\"al\"a foundation. E.~L. thanks the Swedish National Research Council, Vetenskapsr{\aa}det. V.~P. acknowledges the Jenny and Antti Wihuri Foundation for financial support.
\end{acknowledgments}

\bibliography{mqv-manu}
\end{document}